\documentclass[10pt]{article}
\usepackage{amsmath,amsfonts,latexsym}
\usepackage{amssymb}
\usepackage{amscd}
\usepackage[all]{xy}
\advance \textheight by \topskip
\makeatletter

\newtheorem{theorem}{Theorem}[section]

\newtheorem{lemma}[theorem]{Lemma}

\newcommand{\bea}{\begin{eqnarray}}
\newcommand{\eea}{\end{eqnarray}}
\newcommand{\noi}{\noindent}
\newcommand{\e}{\varepsilon}

\newcommand{\tom}{\tilde \Omega}

\begin{document} 
 
\title{The renormalization of the non commutative $\phi_4^{\star 4}$ field theory} 
\author{ 
 Razvan Gurau${}^{(1)}$
\footnote{e-mail:razvan.gurau@th.u-psud.fr}\\ 
${}^{(1)}$Laboratoire de Physique Th\'eorique, B\^at.\ 210, CNRS UMR 8627\\ 
    Universit\'e Paris XI,  F-91405 Orsay Cedex, France
} 
\maketitle 
 
\begin{abstract} 
We present an overview of the different renormalization proofs 
of the non commutative $\phi_4^{\star 4}$ model. This paper is a contribution to the MemPhys project.
\end{abstract} 

\section{Introduction}
\setcounter{equation}{0}

Modern non commutative quantum field theory has been developed over the last twenty years from several reasons. It appears in some limit regime of string theory, so that it might be important for physics beyond the standard model. Predicting the quantization of surface and volume it has also been related to loop quantum gravity.
It is moreover good setup to study the quantum Hall effect.

After  the introduction by Grosse and Wulkenhaar (in \cite{GW1}) of the renormalizable $\phi_4^{\star 4}$ model, it has became clear this theory is interesting {\it per se}: It is much more manageable than string theory or quantum gravity and allows precise mathematical proofs with a very high degree of accuracy.

\section{ The $\phi_4^{\star 4}$ model}
\setcounter{equation}{0}

We will allways take the space Euclidean and flat. We consider the simplest non commutative geometry, the Moyal space. It is defined as the algebra of functions defined on the deformed $\mathbb{R}_{\theta}^4$ space, defined by the non commuting coordinates
\begin{equation}
 [x^\mu, x^\nu]=i \theta^{\mu \nu},  
\end{equation}
where the the matrix $\theta$ is
\begin{eqnarray}
\label{theta}
  \theta= 
  \begin{pmatrix}
      0 &\theta & 0 & 0\\ 
     -\theta & 0 & 0 & 0\\
    0 & 0 & 0 & \theta \\
    0 & 0 & -\theta & 0
    \end{pmatrix}.
\end{eqnarray}
The deformed space $\mathbb{R}_{\theta}^4$ can be seen as the usual  $\mathbb{R}^4$  space endowed with the associative but non commutative Moyal product of functions
\bea
\label{moyal-product} 
 (f\star g)(x)=\frac{1}{\pi^{D}|\det\theta|}\int d^{D}y~d^{D}z~e^{-2\imath y\theta^{-1}z}f(x+y)~g(x+z)\; .
\eea
Using this product we take the action of a scalar field $\phi$ on a Moyal space to be
\bea
\label{lag}
S=\int d^4 x \left(\frac{1}{2} \partial_\mu  \phi
\star \partial^\mu \phi +\frac{\Omega^2}{2} (\tilde{x}_\mu  \phi )\star
(\tilde{x}^\mu \phi ) 
+\frac{\lambda}{4}   \phi \star \phi \star \phi \star \phi \right)
\eea
where  $\tilde{x}_\mu = 2 (\theta^{-1})_{\mu \nu} x^\nu.$ 

This model does not suffer from the well known ultraviolet infrared mixing and as a consequence one has
\begin{theorem}
   The model defined by the action \ref{lag} is perturbatively renormalizable at all orders.
\end{theorem}
Three different proofs of this theorem are presented in the rest of this paper.

\section{Renormalization in the matrix base}
\setcounter{equation}{0}

One can perform a change of basis (see \cite{GW1} for the detailed computations) using Hermite polynomials to cast the action (\ref{lag}) into the (pseudo-) matrix model form
\bea
S=(2\pi\theta)^2\sum_{m,n,k,l\in \mathbb{N}^2}(\phi_{mn}G_{mn,kl}\phi_{kl}+
\frac{\lambda}{4}\phi_{mn}\phi_{nk}\phi_{kl}\phi_{lm}) \, ,
\eea
with 
\bea
G_{mn,kl}&=&2\frac{1+\Omega^2}{\theta}(n^1+n^2+m^1+m^2)\delta_{n^1k^1}\delta_{m^1l^1}
\delta_{n^2k^2}\delta_{m^2l^2}\nonumber\\
&-&2\frac{1-\Omega^2}{\theta}(\sqrt{k^1l^1}\delta_{n^1+1k^1}\delta_{m^1+1l^1}+
\sqrt{m^1n^1}\delta_{m^1-1k^1}\delta_{m^1-1l^1})\delta_{n^2k^2}\delta_{m^2l^2}
-(1\leftrightarrow 2) \, .
\eea
The highly nontrivial computation of the propagator (the inverse of the quadratic part of the action) has been achieved in \cite{GW1}.

For a matrix model, the Feynman graphs are ribbon graphs. They are characterized by some topological numbers, like the number of vertices $N$ (and external vertices $N_e$) the number of lines $L$, the number of internal faces $F$, the genus $g$ and the number of boundaries $B$ of the Riemann surface on which we can draw the graph (see \cite{GW1}). These numbers are related by some topological relations
\bea
N-L+F=2-2g \quad 4N-N_e=2L \, .
\eea

The proof of renormalizability relies on two main results: the counterterm structure (established in \cite{GW1}) and the power counting (established in \cite{RVW}).

The power counting is done using multiscale analysis in the matrix base. The matrix index is (almost) conserved on the internal faces so that we have one main index to sum per internal face. This in term suggests the use of the dual graph (the graph with all vertices replaced by faces and all faces replaced by vertices). Choosing a tree in the dual graph adapted to the scale attribution (\cite{RVW}) one proves that the superficial degree of divergence of a graph is
\bea
\omega(G)=(2-\frac{N_e}{2})-2(2g+B-1)\, .
\eea
Consequently, only $2$ or $4$ point graphs with $g=0$ and $B=1$ can diverge. The divergent part of such graphs is shown to exactly reproduce the terms in the initial action (\cite{GW1}) and can be reabsorbed in a redefinition of the parameters $\Omega$ and $\lambda$.

\section{Renormalization in the direct space}
\setcounter{equation}{0}

The previous proof of renormalizability is not very satisfactory. First of all it is not valid for all values of the parameter $\Omega$. Second is very technical and being performed in the matrix base we lack the intuition to understand the counterterm structure. 

Due to this reasons in \cite{GMRV} an alternative proof is proposed, in the direct space. It is much simpler, provides a principle of ``Moyality'' to replace the principle of locality in commutative field theory and is valid for all values of $\Omega$.

The propagator of the model in the direct space from a point $x$ to a point $y$ is
given by the Mehler kernel (see \cite{GMRV} and references therein)
\begin{equation}
\label{propa1}
C(x,y)=\int_0^\infty \frac{\tilde \Omega d\alpha}{[2\pi\sinh(\alpha)]^{2}}
e^{-\frac{\tilde \Omega}{4}\coth(\frac{\alpha}{2})(x-y)^2-
\frac{\tilde \Omega}{4}\tanh(\frac{\alpha}{2})(x+y)^2}\; ,
\end{equation} 
with $\tilde \Omega=\Omega/(2 \theta)$

Using the explicit form \eqref{moyal-product} 
of the Moyal product, the vertex of of the action \eqref{lag} becomes in the direct space:
\bea
\label{v1}
\delta (x_1^V - x_2^V + x_3^V - x_4^V)e^{2i\sum_{1\le
    i <j\le 4}(-1)^{i+j+1}x_i^V\theta^{-1}x_j^V}
\eea
\noi
where $x_1^V,\ldots, x_4^V$ are the $4-$vectors of the positions of the $4$
fields incident to the respective vertex $V$. The $\delta$ function in the above equation sets the fields in the corners of a paralelogram and the oscillating term is related to the area of the paralelogram. For all lines we change variables to ``short'' $u$ and ``long'' $v$ variables, equal to the sum and difference of its two endpoints.

The sliced propagator is bounded by 
\bea
C^i\le M^{2i}e^{-M^i |u|-M^{i} |v|}\; .
\eea
All vertices are weighted on average by two propagators, thus a $M^{4i}$ factor, and we need to integrate two long and two short variables. We use the $\delta$ function to integrate one of the longs variables. We get a $M^{-8i}$ factor for the short integrations and $M^{4i}$ for the long. Thus the power counting for a vertex is neutral.

This naive argument combined with the use of the oscillating factors provides the power counting (see \cite{GMRV} for the detailed proof). We note that strictu sensu we do not recover the complete power counting as a function of genus and broken faces. We get only a sufficient bound stating that if a graph is non planar or has more than one broken face it is convergent.

To prove the ``Moyality'' of the counterterms, we rewrite the amplitude of a graph using two topological operations, see \cite{GMRV}.
This topological operations are changes of variables adapted to the graph. They are similar to the momentum routing. Thus, using a particular tree in the graph, its oscillating factor can be rewritten in terms of the {\it rosette} of the graph, that is the graph with all tree lines contracted, as

\begin{lemma}
The rosette contribution after a complete first Filk reduction is exactly:
\begin{eqnarray}
&&  \delta(s_1-s_2+\dots-s_{2n+2}+\sum_{l\in T}u_l)
e^{i\sum_{0\leq i<j\leq r}(-1)^{i+j+1}s_i\theta^{-1} s_j}
\nonumber \\
&& e^{-i\sum_{l \prec l'}u_l\theta^{-1} u_{l'}}e^{-i\sum_l \epsilon(l)\frac{u_l\theta^{-1} v_l}{2}} 
e^{i\sum_{l,i \prec l}(-1)^{i} s_i\theta^{-1}u_l+i\sum_{l,i \succ l}u_l \theta^{-1}(-1)^{i} s_i}\ ,
\end{eqnarray}
where $\epsilon(l)$ is $-1$ if the tree line $l$ is oriented towards the root and $+1$ if it is not. 
\end{lemma}

For a planar one broken face graph we have further
\begin{lemma}\label{exactoscill}
The vertex contribution for a planar regular graph is exactly:
\begin{eqnarray}
&&\delta(\sum_{i}(-1)^{i+1}x_{i}+\sum_{l\in T\cup {\cal L}} u_l)
e^{\imath\sum_{i,j}(-1)^{i+j+1}x_{i}\theta^{-1} x_{j}} 
\nonumber \\ 
&&e^{\imath\sum_{l\in T \cup {\cal L},\;   l \prec j}u_l\theta^{-1} (-1)^{j}x_j
+\imath\sum_{l\in T \cup {\cal L},\; l \succ j }(-1)^j x_{j}\theta^{-1} u_l}
\nonumber \\
&&e^{-\imath\sum_{l,l'\in T \cup {\cal L},\; l \prec l' }u_l\theta^{-1} u_{l'}
-\imath\sum_{l\in  T}\frac{u_l\theta^{-1} v_l}{2}\epsilon(l)
-\imath\sum_{l\in {\cal L}}\frac{u_l\theta^{-1} w_l}{2}\epsilon(l)}
\nonumber \\
&&e^{-\imath\sum_{l\in{\cal L},\, l' \in {\cal L} \cup T;\; l'\subset  l}u_{l'}\theta^{-1} w_l \epsilon(l)} \ .
\end{eqnarray}  
\end{lemma}

We are now into the position to understand the ``Moyality'' principle. The divergent part of such a factor is obtained by setting all the shirt $u$ variables to zero. Thus the first two lines of the above lemma hold precisely the form of the Moyal kernel. As such, this divergence can be reabsorbed in the redefinition of the parameters in the initial Lagragian (again, the details are presented in \cite{GMRV}).

\section{Dimensional regularization and renormalization}
\setcounter{equation}{0}

The last renormalization scheme discussed in this paper is the dimensional regularization and renormalization. Although it is not obvious how this procedure can be extended to hold the complete construction of a model, it is by far the most well known as it allows for the perturbative renormalization of gauge theories without braeking the gauge invariance. This procedure relies on the parametric representation of NCQFT introduced in \cite{param1}. Note that for technical reasons we restrict our attention only to the complex model.

We define the $(L\times 4)$-dimensional incidence matrix $\e^V$ for each of the vertices $V$
\bea
\label{r1}
\e_{\ell i}^V= (-1)^{i+1}, \mbox { if the line $\ell$ hooks to the vertex $V$
  at corner $i$.} \nonumber\\
\eta^V_{\ell i}=\vert \e^V_{\ell i}\vert, \mbox { } V=1,\ldots, n,\, 
\ell=1,\ldots, L \mbox{ and } i=1,\ldots, 4. 
\eea
The ''short" $u$ and ''long" $v$ variables are
\bea \label{uv}
v_\ell=\frac{1}{\sqrt{2}} \sum_V \sum_i \eta^V_{\ell i} x^V_i,\quad
u_\ell=\frac{1}{\sqrt{2}} \sum_V \sum_i \e^V_{\ell i} x^V_i.
\eea

Using (\ref{propa1}), (\ref{v1}) and (\ref{uv}) we write the amplitude ${\cal A}_{G,{\bar V}}$ of the graph $G$ (with the marked root $\bar V$) in terms of the non-commutative polynomials $HU_{G, \bar{V}}$ and $HV_{G, \bar{V}}$ as (see \cite{param1} for details)
\bea
\label{HUGV}
{\cal A}_{G,{\bar V}}  (x_e,\;  p_{\bar V}) = \left(\frac{\tom}{2^{\frac D2
      -1}}\right)^L  \int_{0}^{\infty} \prod_{\ell=1}^L  [ d t_\ell
(1-t_\ell^2)^{\frac D2 -1} ]
\frac
{e^{-  \frac {HV_{G, \bar{V}} ( t_\ell , x_e , p_{\bar v})}
{HU_{G, \bar{V}} ( t )}}}
{HU_{G, \bar{V}} ( t )^{\frac D2}},
\eea
with $x_e$ the external positions of the graph and 
$t_\ell = {\rm tanh} \frac{\alpha_\ell}{2}, \ \ell=1,\ldots, L,$
where  $\alpha_\ell$ are the Schwinger parameters of the lines.
We proved in \cite{param1} that  $HU$ and $HV$ are polynomials in the set of variables $t$. 

To write the first polynomial (see \cite{param1}), let $I$ and resp. $J$ be two subsets of $\{1,\ldots,L\}$, of cardinal $\vert I \vert$ and $\vert J \vert$. Furthermore, let $k_{I,J} = \vert I\vert+\vert J\vert - L - F +1 \, ,$
and $n_{I J}$, an integer number (computed in \cite{param1}) and $s=2/(\theta\tilde \Omega)=1/\Omega$. We have
\bea
\label{suma}
HU_{G,{\bar V}} (t) &=&  \sum_{I,J}  s^{2g-k_{I,J}} \ n_{I,J}^2
\prod_{\ell \not\in I} t_\ell \prod_{\ell' \in J} t_{\ell'}\ .
\eea

In \cite{param1}, non-zero {\it leading terms} ({\it i.e.} terms with
the smallest global degree in the $t$ variables) were identified. 
They are dominant in the UV regime. 
Some of them correspond to subsets $I=\{1,\ldots, L\}$ and $J$  {\it admissible}, that is 
\begin{itemize}
\item $J$ contains a tree $\tilde {\cal T}$ in the dual graph,
\item the complement of $J$ contains a tree $\cal T$ in the direct graph.
\end{itemize}
Associated to such $I$ and $J$ one has $n_{I,J}^2=2^{2g}$.

Using this representation the power counting translate in the proof of meromorphy of the function (\ref{HUGV}) in the space-time dimension $D$ (see \cite{Gurau:2007fy} for the detailed proof). In order to prove the meromorhpy one introduces Hepp sectors $\sigma$ defined as 
\bea
\label{hepp-nc}
0\le t_1 \le \ldots \le t_L \, ,
\eea
and performs the change of variables
\bea
\label{change-nc}
t_\ell=\prod_{j=\ell}^L x_j^2,\ \ell =1,\ldots, L.
\eea

We denote by $G_i$ the subgraph composed by the lines $t_1$ to $t_i$. As before, we denote $L(G_i)=i$ the number of lines of $G_i$, $g(G_i)$ its genus, $F(G_i)$ its number of faces, etc..
The amplitude is
\bea
{\cal A}_{G,\bar V}=\Big{(}\frac{\tilde \Omega}{2^{(D-4)/2}}\Big{)}^L
\int_{0}^{1}\prod_{i=1}^L 
\left(1-(\prod_{j=i}^L x_j^2)^2\right)^{\frac D2 -1}
dx_{i} 
\prod_{i=1}^{L}x_{i}^{2L(G_i)-1}
\frac{e^{-\frac{HV_{G,\bar V}(x^2)}{HU_{G,\bar V}(x^2)}}}{HU_{G,\bar V}(x^2)}\, .
\eea 
In the above equation we factor out in $HU_{G,\bar V}$the monomial with the smallest degree in each variable $x_i$
\bea
\label{ampli-x}
{\cal A}_{G,{\bar V}}  (x_e,\;  p_{\bar v}) = \left(\frac{\tom}{2^\frac
    D2}\right)^L  \int_{0}^{1} \prod_{\ell=1}^L  
 dx_\ell \left(1-(\prod_{j=\ell}^L x_j^2)^2\right)^{\frac D2 -1} 
 x_i^{2L(G_i)-1-D b'(G_i)}
 \frac{e^{-\frac {HV_{G, \bar V}}{HU_{G,\bar V}}}}{(a s^b+ F (x^2))^\frac D2}.
\eea
The last term in the above equation is always bounded by a constant. Divergences can arise only in the region $x_i$ close to zero (it is well known that this theory does not have an infrared problem, even at zero mass).
The integer $b'(G_i)$ is given by the topology of $G_i$. It is
\bea
b'(G_i)=\begin{cases}
    {\displaystyle \le L(G_i)-[n(G_i)-1]-2g(G_i)} &\text{if } 
      g(G_i)>0
     \vspace{.3cm}\\
     {\displaystyle \le L(G_i)-n(G_i)} &\text{if }
      g(G_i)=0 \text{ and } B(G_i)>1
     \vspace{.3cm}\\
     {\displaystyle =L(G_i)-[n(G_i)-1]} &\text{if }
       g(G_i)=0 \text{ and } B(G_i)=1 \\
  \end{cases} .
\eea
To prove this formula one uses  the leading terms in the polynomial $HU$. Taking individually the integral over each $x_i$ we see that only planar one broken face subgraphs with two or four external legs are divergent. 

In this context the Moyality is replaced by the following factorization property (proved in \cite{Gurau:2007fy})
\bea
\label{eq:factfinal}
\frac{e^{-\frac{HV_G(\rho)}{HU_G(\rho)}}}{HU_G(\rho)^{D/2}}=
\frac{1}{[HU^{l(\rho)}_{S}]^{D/2}}(1+\rho^2{\cal O}_S) \frac{e^{-\frac{HV_{G/S}}{HU_{G/S}}}}{HU_{G/S}^{D/2}} \, .
\eea
of the amplitude of a graph under rescaling by $\rho$ of the parameters of a subgraph. The pole subtraction is done by the usual Taylor operator (again see \cite{Gurau:2007fy} for details).

\section{Conclusions}

Renormalizable non commutative quantum field theories are today a well established field. Most of the techniques of commutative field theories can be generalized to RNCQFT. The main renormalization proofs of the latter have been presented in this paper. The structure of RNCQFT is more topological than that of usual QFT's and might be more adapted to the analysis of background independent theories.

\end{document}